\documentclass[11pt,showpacs]{revtex4}
\def\comment#1{}

\usepackage{graphicx}
\begin{document}
\title{Electrodynamics for Nuclear Matter in Bulk}
\author{Remo Ruffini, Michael Rotondo and She-Sheng Xue}
\email{ruffini@icra.it ; Michael.Rotondo@icra.it  ; xue@icra.it}
\affiliation{ICRANet and
Physics Department, University of Rome ``La Sapienza", 00185 Rome, Italy}
\begin{abstract}
A general approach to analyze the electrodynamics of nuclear matter in bulk is presented using  the relativistic Thomas-Fermi equation generalizing to the case of $N \simeq (m_{\rm Planck}/m_n)^3$ nucleons of mass $m_n$ the approach well tested in very heavy nuclei ($Z \simeq 10^6$). Particular attention is given to implement the condition of charge neutrality globally on the entire configuration, versus the one usually adopted on a microscopic scale. As the limit $N \simeq (m_{\rm Planck}/m_n)^3$ is approached the penetration of electrons inside the core increases and a relatively small tail of electrons persists leading to a significant electron density outside the core. Within a region of 
$10^2$ electron Compton wavelength near the core surface electric fields close to the critical value for pair creation by vacuum polarization effect develop. These results can have important consequences on the understanding of physical process in neutron stars structures as well as on the initial conditions leading to the process of gravitational collapse to a black hole.
\end{abstract}
\maketitle

It is well know that the Thomas-Fermi equation is the exact theory for atoms, molecules and solids as $Z\rightarrow\infty$ \cite{liebsimon}. We show in this letter that the relativistic Thomas-Fermi theory developed for the study of atoms for heavy nuclei with $Z \simeq 10^6$ \cite{pomer1945}-\cite{ruffinistella81} gives important basic new information on the study of nuclear matter in bulk in the limit of $N \simeq (m_{\rm Planck}/m_n)^3$ nucleons of mass $m_n$ and on its electrodynamic properties. 
The analysis of nuclear matter bulk in neutron stars composed of degenerate gas of neutrons, protons and electrons, 
has traditionally been approached by implementing microscopically the charge neutrality condition  by requiring the electron density $n_e(x)$ to coincide with the proton density $n_p(x)$,
\begin{eqnarray}
n_e(x)=n_p(x).
\label{localnp}
\end{eqnarray}
It is clear however that especially when conditions close to the gravitational collapse occur, there is an ultra-relativistic component of degenerate electrons whose confinement requires the existence of very strong 
electromagnetic fields, in order to guarantee the overall charge neutrality of the neutron star. Under these conditions equation (\ref{localnp}) will be necessarily violated. We are going to show in this letter that they will develop electric fields close to the critical value $E_c$ introduced by Sauter \cite{sauter}, Heisenberg and Euler \cite{euler}, and by 
Schwinger \cite{schwinger} 
\begin{eqnarray}
E_c=\frac{m^2c^3}{e\hbar}.
\label{ec}
\end{eqnarray}
Special attention for the existence of critical electric fields and the possible condition for electron-positron ($e^+e^-$)
pair creation out of the vacuum  in the case of heavy bare nuclei, with the atomic number $Z\geq 173$, has been 
given by Pomeranchuk and Smorodinsky \cite{pomer1945}, Gershtein and Zel'dovich \cite{zeldo70}, 
Popov \cite{popov}, Popov and Zel'dovich \cite{zeldovich}, 
Greenberg and Greiner \cite{greiner82}, Muller, Peitz, Rafelski and Greiner \cite{muller72}.
They analyzed the specific pair creation process of an electron-positron pair around both a point-like 
and extended bare nucleus by direct integration of Dirac equation.
These considerations have been extrapolated to much heavier nuclei $Z\gg 1600$, implying the creation of a large number of  $e^+e^-$ pairs, by using a statistical approach based on the relativistic Thomas-Fermi equation by Muller and Rafelski \cite{muller75}, Migdal, Voskresenskii 
and Popov \cite{migdal76}. 
Using substantially the same statistical approach based on the relativistic Thomas-Fermi equation, Ferreirinho et al. \cite{ruffinistella80}, Ruffini and Stella \cite{ruffinistella81} have analyzed the electron densities around an extended nucleus in a neutral 
atom all the way  up to $Z\simeq 6000$. They have shown the effect of penetration of the electron orbitals well 
inside the nucleus, leading to a screening of the nuclei positive charge and to the concept of an ``effective'' nuclear charge distribution.
All the above works assumed for the radius of the extended nucleus the semi-empirical formulae \cite{segrebook},
\begin{eqnarray}
R_c\approx r_0 A^{1/3},\quad r_0=1.2\cdot 10^{-13}{\rm cm},
\label{dn}
\end{eqnarray}
where the mass number $A=N_n+N_p$, $N_n$ and $N_p$ are the neutron and proton numbers.  
The approximate relation between $A$ and the atomic number $Z=N_p$, 
\begin{eqnarray}
Z \simeq \frac{A}{2},
\label{z2a}
\end{eqnarray}
was adopted in Refs. \cite{muller75,migdal76}, or the empirical formulae
\begin{eqnarray}
Z &\simeq & [\frac{2}{A}+\frac{3}{200}\frac{1}{A^{1/3}}]^{-1},
\label{zae}
\end{eqnarray}
was adopted in Refs. \cite{ruffinistella80,ruffinistella81}.
 
\begin{figure}[th] 
\begin{center}
\includegraphics[width=10.5cm,height=8.5cm]{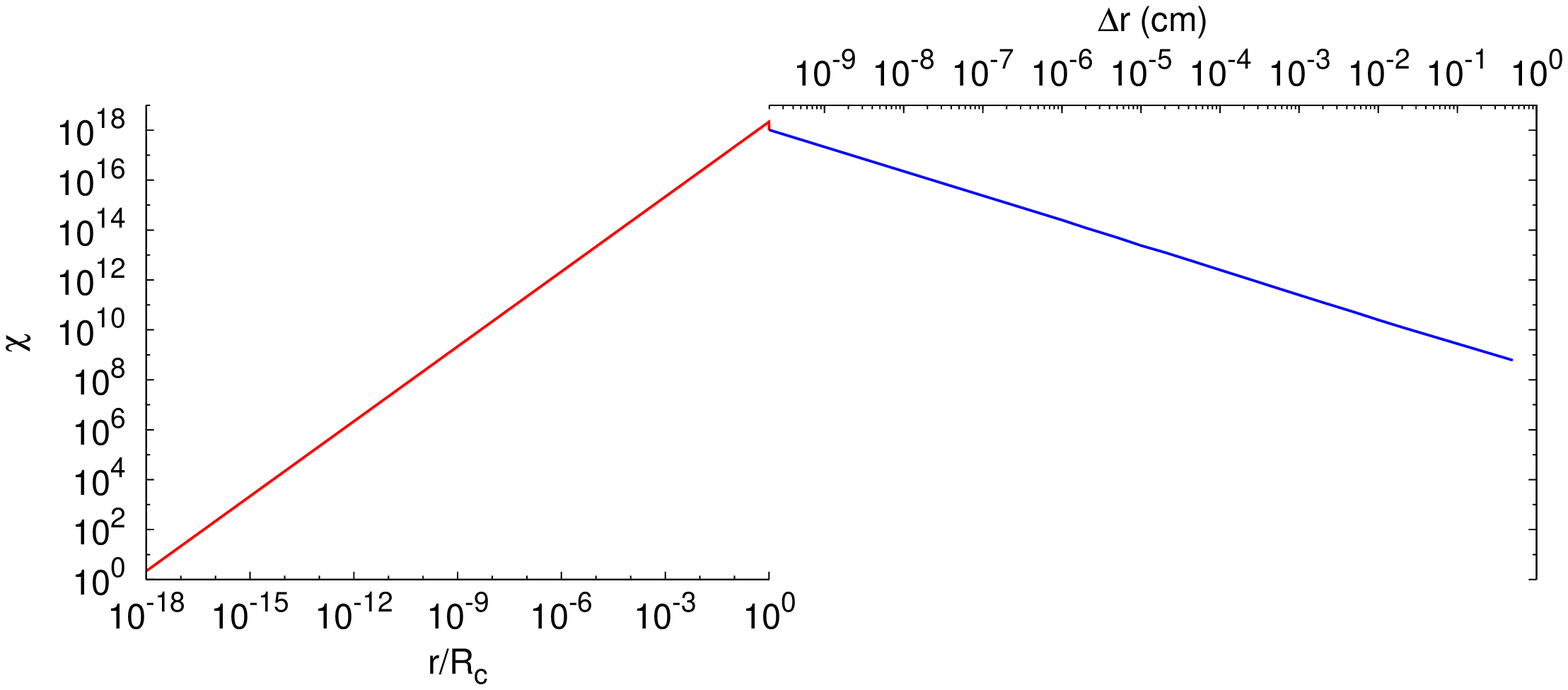}
\end{center}
\caption{The solution $\chi$ of the relativistic Thomas-Fermi Equation for $A=10^{57}$ and core radius $R_c=10$km, is plotted 
as a function of radial coordinate. The left red line corresponds to the internal solution and it is plotted as a function of radial coordinate in unit of $R_c$ in logarithmic scale. The right blue line corresponds to the solution external to the core and it is plotted as function of the distance $\Delta r$ from the surface in the logarithmic scale in centimeter.}%
\label{chif}%
\end{figure}

\begin{figure}[th]
\begin{center}
\includegraphics[width=10.5cm,height=8.5cm]{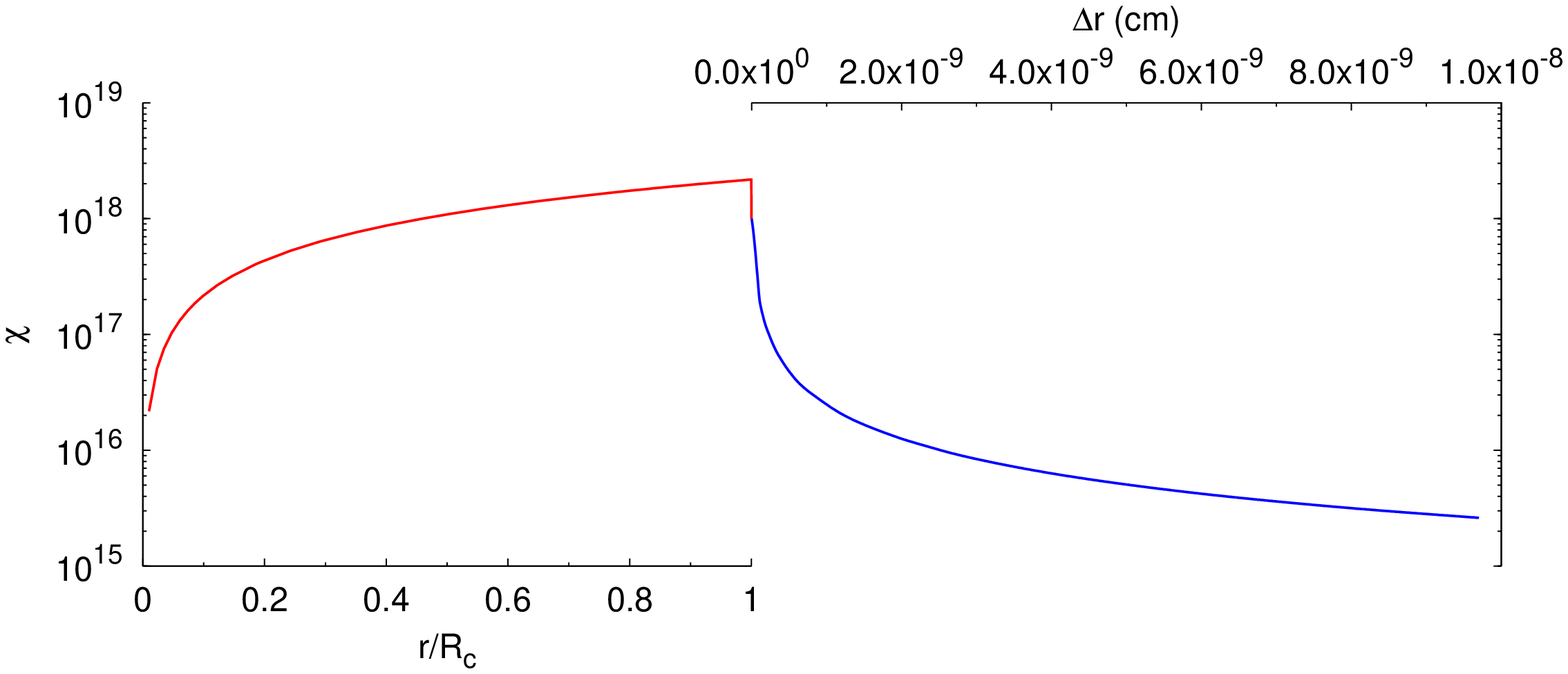}
\end{center}
\caption{The same as Fig.~(\ref{chif}): enlargement around the core radius $R_c$ showing explicitly the continuity of function $\chi$ and its 
derivative $\chi'$ from the internal to the external solution. }%
\label{chircf}%
\end{figure}

The aim of this letter is to outline an alternative approach of the description of nuclear matter in bulk: it generalizes, to the case of $N \simeq (m_{\rm Planck}/m_n)^3$ nucleons, the above treatments, already developed and tested for the study of heavy nuclei. 
This more general approach differs in many aspects from the ones in the current literature and recovers, in the limiting case of $A$ 
smaller than $10^6$, the above treatments. We shall look for a solution implementing the condition of overall charge neutrality of the star as given by 
\begin{eqnarray}
N_e=N_p,
\label{golbalnp}
\end{eqnarray}
which significantly modifies Eq.~(\ref{localnp}), since now $N_e (N_p)$ is the total number of electrons (protons) of the equilibrium configuration.
Here we present only a
simplified prototype of this approach. We outline the essential relative role
of the four fundamental interactions present in the neutron star physics: the gravitational, weak, strong and
electromagnetic interactions. In addition, we also implement the fundamental role of Fermi-Dirac statistics   
and the phase space blocking due to the Pauli principle in the degenerate configuration. The new results essentially depend from the coordinated action of the five above theoretical components and cannot be obtained if any one of them is neglected.
Let us first recall the role of gravity.
In the case of neutron stars, unlike in the case of nuclei where its effects can be neglected, gravitation has the fundamental role of defining the  basic parameters of the equilibrium configuration. As pointed out by Gamow \cite{gamow-book}, at a Newtonian level and by Oppenheimer and Volkoff \cite{OV39} in general relativity, configurations of equilibrium exist at approximately one solar mass and at an average density around the nuclear density. This result is obtainable considering only the gravitational interaction of a system of Fermi degenerate self-gravitating neutrons, neglecting all other particles and interactions. It can be formulated within a Thomas-Fermi self-gravitating model 
(see e.g. \cite{ruffiniphd}). 
In the present case of our simplified prototype model directed at evidencing new electrodynamic properties, the role of gravity is simply taken into account by considering, in line with the generalization of the above results, a mass-radius relation for the baryonic core
\begin{eqnarray}
R^{NS}=R_c\approx \frac{\hbar}{m_\pi c}\frac{m_{\rm Planck}}{m_n} .
\label{dnns}
\end{eqnarray}
This formula generalizes the one given by Eq.~(\ref{dn}) extending its validity  to  $N\approx (m_{\rm Planck}/m_n)^3$, 
leading to a baryonic core radius $R_c\approx  10$km.
We also recall that a more detailed analysis of nuclear matter in bulk in neutron stars ( see e.g. Bethe et al. \cite{sato1970} and Cameron \cite{cameron1970} ) shows that at mass densities larger than the "melting" density of 
\begin{eqnarray}
\rho_c=4.34 \cdot 10^{13} g/cm^3,
\label{melting}
\end{eqnarray}
all nuclei disappear. In the description of nuclear matter in bulk we have to consider then the three Fermi degenerate gas of neutrons, protons and electrons. In turn this naturally leads to consider the role of strong and weak interactions among the nucleons. In the nucleus, the role of the strong and weak interaction, with a short range of one Fermi, is to bind the nucleons, with a binding energy of 8 MeV, in order to balance the Coulomb repulsion of the protons. In the neutron star case we have seen that the neutrons confinement is due to gravity. We still assume that an essential role of the strong interactions is to balance the effective Coulomb repulsion due to the protons, partly screened by the electrons distribution inside the neutron star core. We shall verify, for self-consistency, the validity of this assumption on the final equilibrium solution we are going to obtain.
We now turn to the essential weak interaction role in establishing the relative balance between neutrons, protons and electrons via the direct and inverse 
$\beta$-decay
\begin{eqnarray}
p+ e  &\longrightarrow & n + \nu_e ,
\label{beta}\\
n  &\longrightarrow & p +e + \bar\nu_e.
\label{ibeta}
\end{eqnarray} 
Since neutrinos escape from the star and the Fermi energy of the electrons is null, as we will show below, the only non-vanishing terms in the equilibrium condition given by the weak interactions are: 
\begin{eqnarray}
[(P_n^Fc)^2+M^2_nc^4]^{1/2}-M_nc^2=  [(P_p^Fc)^2+M^2_pc^4]^{1/2}-M_pc^2  + |e|V^p_{\rm coul},  
\label{neq}
\end{eqnarray}
where $P_n^F$ and $P_p^F$  are respectively, the neutron and proton Fermi momenta, and $V^p_{\rm coul}$ is the Coulomb potential of protons. At this point, having fixed all these physical constraints, the main task is to find the electrons distributions fulfilling in addition to the Dirac-Fermi statistics also the Maxwell equations for the electrostatic. The condition of equilibrium of the  Fermi degenerate electrons implies the null value of the Fermi energy:
\begin{eqnarray}
[(P_e^Fc)^2+m^2c^4]^{1/2}-mc^2  + eV_{\rm coul}(r)=0,
\label{eeq}
\end{eqnarray}
where $P_e^F$ is the electron Fermi momentum and $V_{\rm coul}(r)$ the Coulomb potential.
In line with the procedure already followed for the heavy atoms  
\cite{ruffinistella80},\cite{ruffinistella81} we here adopt the relativistic Thomas-Fermi Equation:
\begin{eqnarray}
\frac {1}{x}\frac {d^2\chi(x)}{dx^2}= - 4\pi \alpha\left\{\theta(x-x_c)
- \frac {1}{3\pi^2}\left[\left(\frac {\chi(x)}{x}+\beta\right)^2-\beta^2\right]^{3/2}\right\},
\label{eqless}
\end{eqnarray}
where $\alpha=e^2/(\hbar c)$, $\theta(x-x_c)$ represents the normalized proton density distribution, the variables $x$  and $\chi$  are related to the radial coordinate and the electron Coulomb potential $V_{\rm coul}$ by 
\begin{eqnarray}
x=\frac {r}{R_c}\left(\frac {3N_p}{4\pi}\right)^{1/3};\quad eV_{\rm coul}(r)\equiv \frac {\chi(r)}{r},
\label{dless}
\end{eqnarray} 
and the constants $x_c (r=R_c)$ and $\beta$ are respectively
\begin{eqnarray}
x_c\equiv\left(\frac {3N_p}{4\pi}\right)^{1/3};\quad \beta\equiv  
\frac {mcR_c}{\hbar}\left(\frac{4\pi}{3N_p}\right)^{1/3}.
\label{dbeta}
\end{eqnarray}
The solution has the boundary conditions
\begin{eqnarray}
\chi(0)=0;\quad \chi(\infty)=0,
\label{bchi}
\end{eqnarray} 
with the continuity of the function $\chi$ and its first derivative $\chi'$ at the boundary of the core $R_c$.
The crucial point is the determination of the eigenvalue of the first derivative at the center 
\begin{eqnarray}
\chi'(0)={\rm const}. ,
\label{bchi1}
\end{eqnarray} 
which has to be determined by fulfilling the above boundary conditions (\ref{bchi}) and constraints given by 
Eq.~(\ref{neq}) and Eq.~(\ref{golbalnp}).
The difficulty of the integration of the Thomas-Fermi Equations is certainly one of the most celebrated chapters in theoretical physics and mathematical physics, still challenging a proof of the existence and uniqueness of the solution and strenuously avoiding the occurrence of exact analytic solutions. We recall after the original papers of Thomas \cite{Thomas} and Fermi \cite{Fermi}, the works of Scorza Dragoni \cite{scorza}, Sommerfeld \cite{sommerfeld}, Miranda \cite{Miranda}  all the way to the many hundredth papers reviewed in the classical articles of Lieb and Simon \cite{liebsimon}, Lieb \cite{Lieb} and Spruch \cite{Spruch}.  The situation here is more difficult since we are working on the special relativistic generalization of the Thomas-Fermi Equation. 
Also in this case, therefore, we have to proceed by numerical integration. The difficulty of this numerical task is further enhanced by a consistency check in order to fulfill all different constraints. It is so that we start the computations by assuming a total number of protons and a value of the core radius $R_c$. We integrate the Thomas-Fermi Equation and we determine the number of neutrons from the Eq.~(\ref{neq}). We iterate the procedure until a value of $A$ is reached consistent 
with our choice of the core radius. The paramount difficulty of the problem is the numerical determination of the 
eigenvalue in Eq.~(\ref{bchi1}) which already for $A \approx 10^{4}$ had presented remarkable numerical difficulties \cite{ruffinistella80}. In the present context we have been faced for a few months by an  apparently unsurmountable numerical task: the determination of the eigenvalue seemed to necessitate a  significant number of decimals in the first derivative (\ref{bchi1}) comparable to the number of the electrons in the problem! 
We shall discuss elsewhere the way we overcame the difficulty by splitting the problem on the ground of the physical interpretation of the solution \cite{rrxtbp}. The solution is given in 
Fig.~(\ref{chif}) and Fig.~(\ref{chircf}).

\begin{figure}[th]
\begin{center}
\includegraphics[width=10.5cm,height=8.5cm]{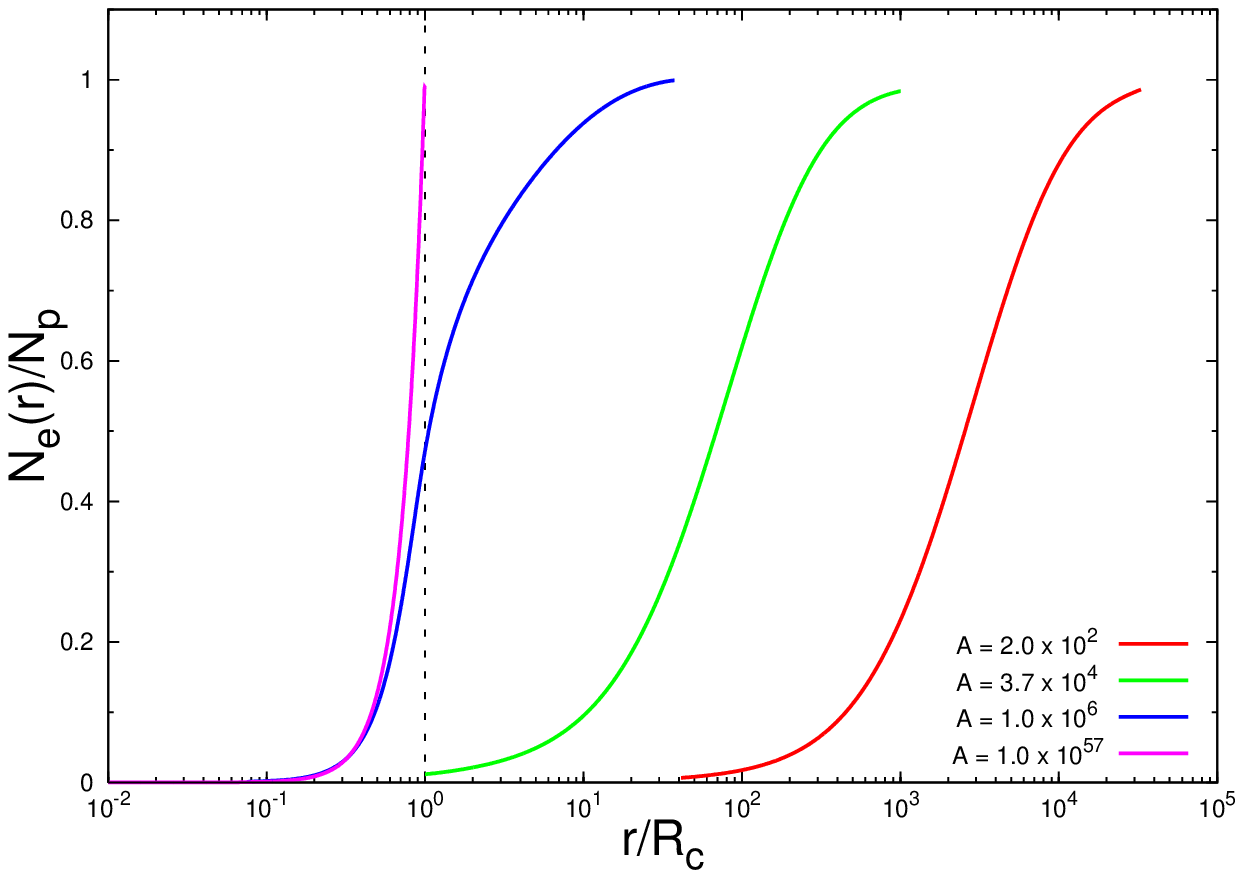}
\end{center}
\caption{The electron number (\ref{tein}) in the unit of the total proton number $N_p$, for selected values of 
$A$, is given as function of radial distance in the unit of the core radius $R_c$, again in logarithmic scale. It is clear how by increasing the value of $A$ the penetration of electrons inside the core increases. The detail shown in Fig.~(\ref{efieldf}) and Fig.~(\ref{enumberf1}) demonstrates how  for $N \simeq (m_{\rm Planck}/m_n)^3$ a relatively small
tail of electron outside the core exists and generates on the baryonic core surface an electric field close to the critical value given in . A significant electron density outside the core is found.
}%
\label{enumberf}%
\end{figure}

\begin{figure}[th]
\begin{center}
\includegraphics[width=10.5cm,height=8.5cm]{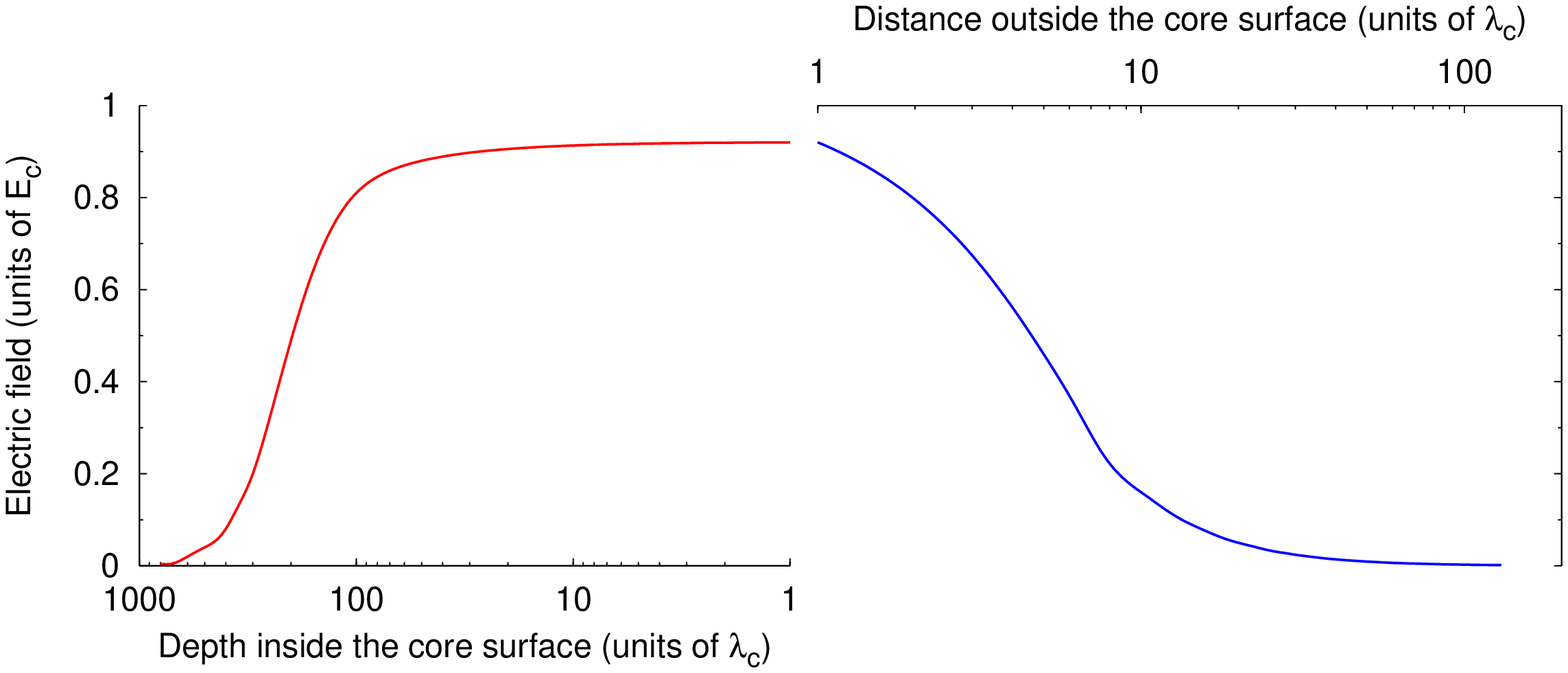}
\end{center}
\caption{The electric field in the unit of the critical field $E_c$ is plotted around the core radius $R_c$. The left (right) diagram in the red (blue) refers the region just inside (outside) the core radius plotted logarithmically. By increasing the density of the star the field approaches the critical field. }%
\label{efieldf}%
\end{figure}

\begin{figure}[th]
\begin{center}
\includegraphics[width=10.5cm,height=8.5cm]{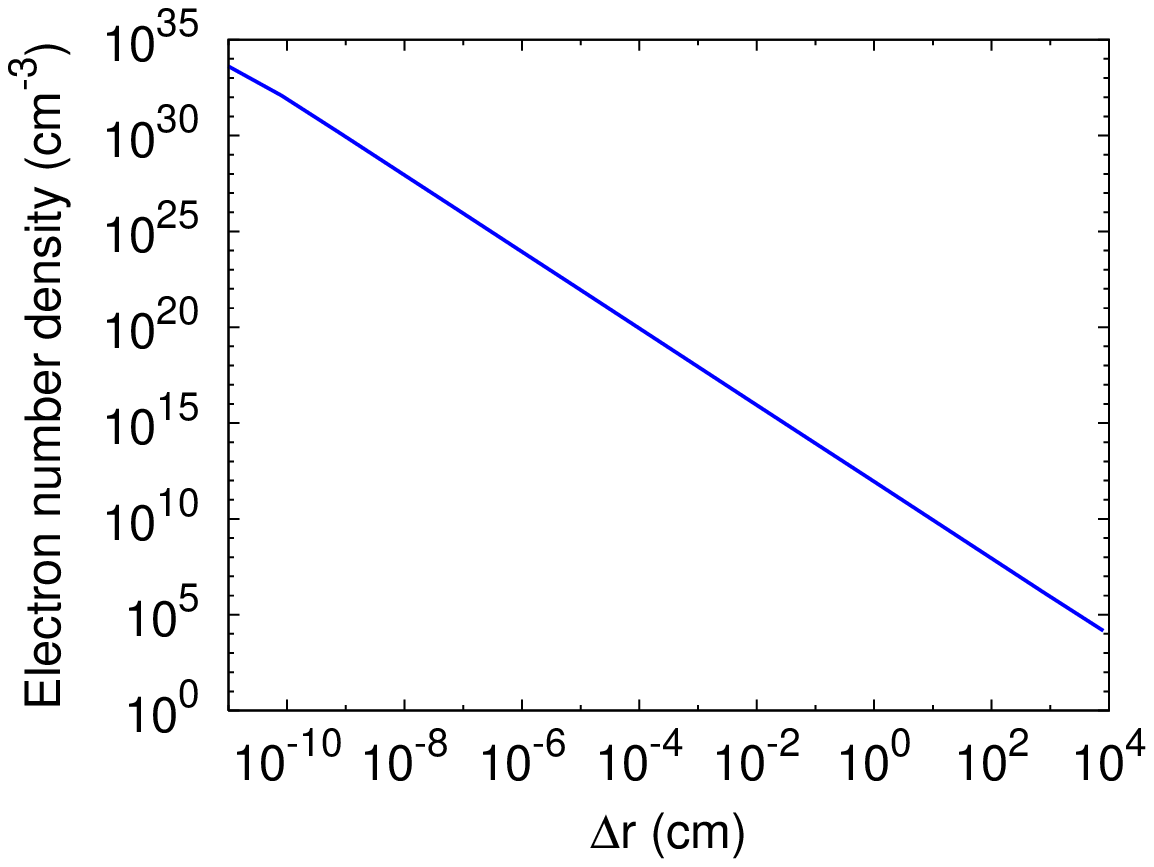}
\end{center}
\caption{ The density of electrons for $A=10^{57}$ in the region outside the core; both scale are logarithmically.
}%
\label{enumberf1}%
\end{figure}

A relevant quantity for exploring the physical significance of the solution is given by the number of electrons within a given radius $r$:
\begin{eqnarray}
N_e(r)=\int_0^{r} 4\pi (r')^2 n_e(r')dr'.
\label{tein}
\end{eqnarray}
This allows to determine, for selected values of the $A$ parameter, the distribution of the electrons within and outside the core and follow the progressive penetration of the electrons in the core at increasing values of $A$ [ see Fig.~(\ref{enumberf})]. 
We can then evaluate, generalizing the results in \cite{ruffinistella80}, \cite{ruffinistella81} , the net charge inside the core
\begin{eqnarray}
N_{\rm net} = N_p-N_e(R_c) < N_p,
\label{net}
\end{eqnarray} 
and consequently determine of the electric field at the core surface, as well as within and outside the core 
[see Fig.~(\ref{efieldf})] and evaluate as well the  Fermi degenerate electron distribution outside the core 
[see Fig.~(\ref{enumberf1})].
It is interesting to explore the solution of the problem under the same conditions and constraints imposed by the fundamental interactions and the quantum statistics and imposing instead of Eq.~(\ref{localnp}) 
the corresponding Eq.~(\ref{golbalnp}). Indeed a solution exist and is much simpler  
\begin{eqnarray}
n_n(x)=n_p(x)=n_e(x)=0,\quad \chi=0.
\label{trivial}
\end{eqnarray}


Before concluding as we announce we like to check on the theoretical consistency of the solution. We obtain an overall neutral configuration for the nuclear matter in bulk, with a positively charged baryonic core with  
\begin{equation}
N_{\rm net}= 0.92\left(\frac{m}{m_\pi}\right)^2\left(\frac{e}{m_n\sqrt{G}}\right)^2\left(\frac{1}{\alpha}\right)^2 ,
\label{nete}
\end{equation}
and an electric field on the baryonic core surface (see Fig.~(\ref{efieldf}) )
\begin{equation}
\frac{E}{E_c}=0.92.
\label{esurface}
\end{equation} 
The corresponding Coulomb repulsive energy per nucleon is given by 
\begin{equation}
U^{\rm max}_{\rm coul}= \frac{1}{2\alpha}\left(\frac{m}{m_\pi}\right)^3mc^2\approx 1.78\cdot 10^{-6}({\rm MeV}),
\label{coul1}
\end{equation}
well below the nucleon binding energy per nucleon. It is also important to verify that this charge core is gravitationally stable.
We have in fact 
\begin{equation}
\frac{Q}{\sqrt{G}M}=\alpha^{-1/2}\left(\frac{m}{m_\pi}\right)^2\approx 1.56\cdot 10^{-4}.
\label{nuclb}
\end{equation}
The electric field of the baryonic core is screened to infinity by an electron distribution given in Fig.~(\ref{enumberf1}).
As usual any new solution of Thomas-Fermi systems has relevance and finds its justification in the theoretical physics and mathematical physics domain. We expect that as in the other solutions previously obtained in the literature of the relativistic Thomas-Fermi equations also this one we present in this letter will find important applications in physics and astrophysics.
There are a variety of new effects that such a generalized approach naturally leads to: (1) the mass-radius relation 
of neutron star may be affected; (2) the electrodynamic aspects of neutron stars and pulsars will be 
different; (3) we expect also important 
consequence in the initial conditions in the physics of gravitational collapse of the baryonic core as soon as the critical mass for gravitational collapse to a black hole is reached. The consequent collapse to a black hole will have very different energetics properties.

\end{document}